\documentclass{Interspeech}

\interspeechcameraready

\title{Discrete Tokens Exhibit Interlanguage Speech Intelligibility Benefit:
an Analytical Study Towards Accent-robust ASR
Only with Native Speech Data}

\author[affiliation={1,2}]{Kentaro}{Onda}
\author[affiliation={3}]{Keisuke}{Imoto}
\author[affiliation={2}]{Satoru}{Fukayama}
\author[affiliation={1}]{Daisuke}{Saito}
\author[affiliation={1}]{Nobuaki}{Minematsu}

\affiliation{}{The University of Tokyo}{Japan}
\affiliation{}{National Institute of Advanced Industrial Science and Technology (AIST)}{Japan}
\affiliation{}{Kyoto University}{Japan}
\email{\{ondakentaro, mine\}@gavo.t.u-tokyo.ac.jp}

\usepackage{comment}
\usepackage{url}
\usepackage{multirow}
\usepackage{arydshln}
\usepackage{caption}
\usepackage{booktabs}
\usepackage{cellspace}
\usepackage{array}
\usepackage[dvipdfmx]{graphicx}

\begin{document}

\maketitle

\begin{abstract}
  In this study, we gained insight that contributes to 
  achieving accent-robust ASR using only native speech data. 
  In human perception of non-native speech, the phenomenon 
  known as ``interlanguage speech intelligibility beneﬁt" (ISIB) is observed, 
  where non-native listeners who share the native language with the speaker 
  understand the speech better compared even to native listeners. 
  Based on the idea that discrete tokens extracted from 
  self-supervised learning (SSL) models represent the human perception 
  of speech, we conducted an analytical study on the robustness of 
  discrete token-based ASR to non-native speech, 
  varying the language used for training the tokenization, 
  which is viewed as a technical implementation of ISIB. 
  The results showed that ISIB actually occurred in 
  the discrete token-based ASR. Since our approach relies only on 
  native speech data to simulate the behavior of human perception, 
  it is expected to be applicable to a wide range of accents for which speech data 
  is scarce.
\end{abstract}

\section{Introduction}
Automatic speech recognition (ASR) is widely used in applications such as 
voice-controlled devices and transcription. 
ASR systems are required to be robust against audio variations caused by
factors such as speakers and recording environments.
However, it is well known that ASR performance tends to degrade
when recognition is performed on accented speech by non-native speakers \cite{racial}.
In today's increasingly globalized world, communication between people 
with different native languages is becoming more common. 
Furthermore, in recent language education, the ``intelligibility principle" has 
become dominant, which emphasizes that learners' pronunciation does not need to be native-like as long as 
it is intelligible enough \cite{munro1995foreign, murphy2014intelligible, levis2020revisiting}. 
As a result, people are more likely to speak with their own accents,
and the demand for ASR systems capable of recognizing 
foreign-accented speech has grown even further.

To achieve accent-robust ASR, various methods such as 
data augmentation and accent adaptation have been proposed \cite{fukuda18_interspeech,Klumpp2023SyntheticCD,prabhu-etal-2023-accented,prabhu24b_interspeech}. 
However, these methods generally require 
accented speech data for training, limiting their applicability
only to a few types of accents for which such corpora are available. 
They are mostly of ``X-accented English" \cite{hinsvark2021accentedspeechrecognitionsurvey}, 
while there are much more diverse combinations of 
``X-accented Y" spoken in the real world.
On the other hand, native speech data, both for language X and Y, 
are relatively easier to obtain for various languages.
If these data are used effectively to achieve accent-robust ASR,
such approach could be applicable to a wide range of language accents,
including those for which accented speech data are scarce or even unavailable.

In \cite{onda24_interspeech}, a method was proposed to synthesize foreign-accented speech using only native speech data. 
This approach assumes the resynthesis of speech using discrete tokens extracted from
self-supervised learning (SSL) models 
as a model of human 
speech reproduction, where speech is perceived as discrete symbols, which are then vocalized.
By resynthesizing the input speech 
with a model trained on a different language,
the model can generate speech that sounds like the reproduction
of the input speech by a non-native speaker.
This method is based on the hypothesis that the process of
discretization simulates
how humans with different native languages 
perceive speech. 
For example, discrete tokens trained on Japanese speech
are assumed to 
represent the results of the speech perception experienced
by native Japanese speakers.

In the context of human perception of non-native speech,
a phenomenon known as ``interlanguage speech intelligibility benefit" (ISIB)
is often observed \cite{bent2003interlanguage,harding2012accent,XIE2013369}.
This refers to the tendency where non-native speech is more intelligible
to listeners who share the same native language with the speaker
than to native listeners of the spoken language.
For example,
Japanese listeners tend to have an advantage in 
understanding Japanese-accented English compared to native English listeners.

In this study, as a first step towards accent-robust ASR using only native speech data,
we investigated whether ISIB can be observed in discrete token-based ASR.
Specifically, we examined whether the recognition performance for foreign-accented speech
can be improved by using discrete tokens trained on the speaker's native language.
This could also validate the hypothesis that
discrete tokens indeed represent the human perception of speech.
Unlike synthesis task performed in \cite{onda24_interspeech},
where acoustic information is added to the tokens
during vocalization process, the output of ASR consists purely of linguistic information.
This makes it possible to conduct the validation
in a more interpretable way.

Our scientific contributions are summarized as follows:
\begin{itemize}
    \item We conducted an analytical study on the robustness of discrete token-based ASR to foreign-accented speech,
    changing the language used for training the tokenization as a technical implementation of ISIB. 
    \item We confirmed the occurrence of ISIB in the discrete token-based ASR, 
    which provides additional support for the hypothesis that discrete tokens represent the human perception of speech.
    \item We suggested the potential of our approach, using discrete tokens to simulate ISIB,
    for achieving accent-robust ASR using only native speech data.
\end{itemize}


\section{Related studies}
\subsection{Generative Spoken Language Model}
Generative Spoken Language Model (GSLM) \cite{lakhotiaetal2021generative} is a language model trained only 
with speech data without any text.
It was proposed under the concept of ``textless NLP,'' based on the idea that
text is not always necessary for human language acquisition.
In GSLM, ``pseudo-text" is used as a substitute for text, which is learned
through k-means clustering of features exracted from 
SSL models such as HuBERT \cite{hubert} and wav2vec2.0 \cite{wav2vec}.
This pseudo-text, also referred to as ``units'' or ``discrete tokens,''
has been explored for various applications across different tasks \cite{chang24b_interspeech}.
Our experiment also utilizes this discrete tokens as intermediate representations
for ASR.

\subsection{Simulation of foreign accentuation only with native speech data}
\label{subsec:simulation}

\begin{figure}[t]
  \centering
  \hspace*{-6mm}
  \includegraphics[width=1.1\linewidth]{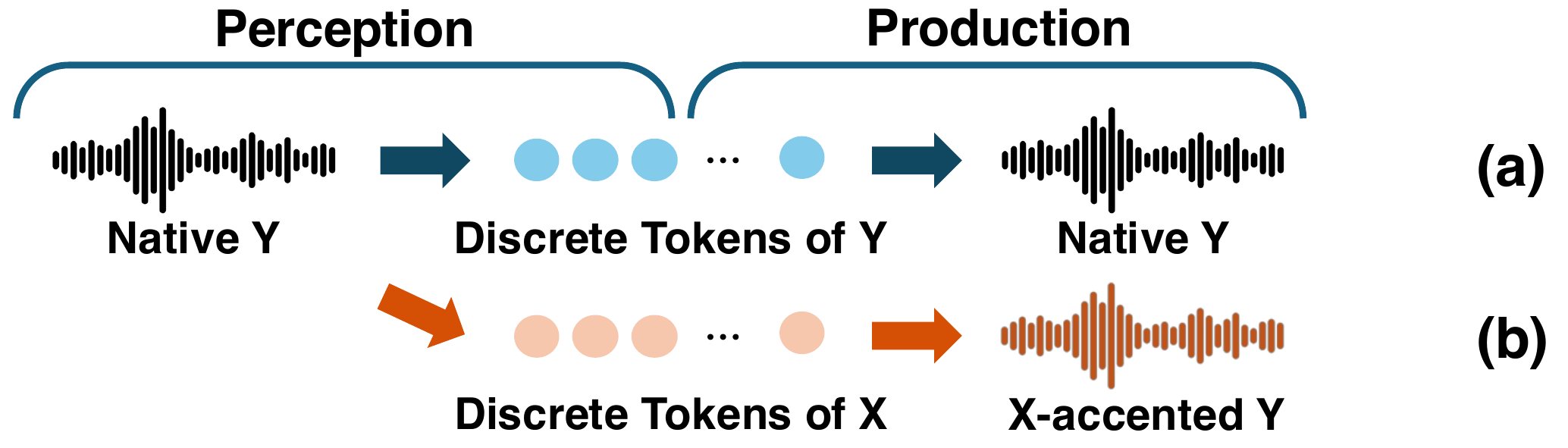}
  \vspace*{-5mm}
  \caption{The reproduction of native speech of language Y by (a) native speakers of language Y, (b) native speakers of language X: the output speech of (b) is expeted to be ``X-accented Y".}
  \label{fig:resynth}
  \vspace*{-2mm}
\end{figure}

\begin{figure}[t]
  \centering
  \hspace*{-6mm}
  \vspace*{-2mm}
  \includegraphics[width=1.1\linewidth]{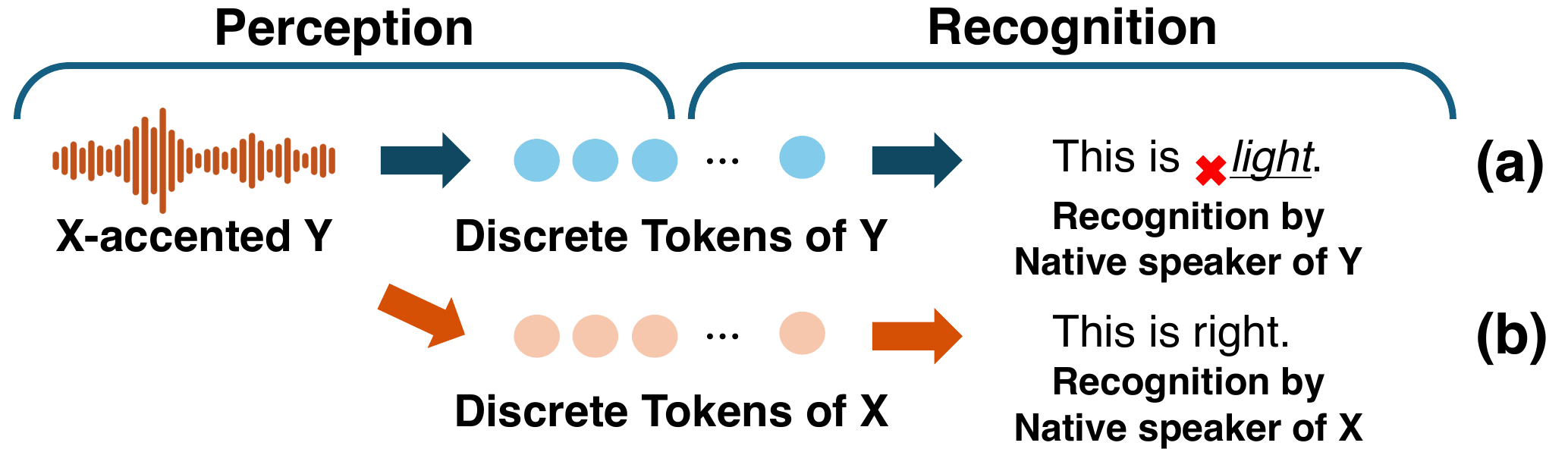}
  \vspace*{-3mm}
  \caption{The recognition of ``X-accented Y" speech by (a) native speakers of language Y, (b) native speakers of language X: (b) is expected to be more accurate through the ISIB effect.}
  \label{fig:transcription}
  \vspace*{-4mm}
\end{figure}
In \cite{onda24_interspeech}, a method for synthesizing foreign-accented speech
using only native speech data was proposed.
This method is based on the idea that the discrete tokens
in GSLM represents the human perception of speech.
Foreign accents occur
when speech in a foreign language is perceived
within the framework of the listener's native language.
When the listener repeats it, 
the perceived speech is vocalized
in a way also influenced by the native language \cite{flege1995second}.
This human process of speech reproduction was modeled as speech resynthesis using discrete tokens
as intermediate representations, as shown in Figure \ref{fig:resynth}.
By training the tokenization and its vocalization with native speech data of language X,
which is different from the input speech of language Y,
it is possible to generate ``X-accented Y'' speech.
Our study extends this idea of regarding discrete tokenization as a model of human perception
to ASR for foreign-accented speech.

\subsection{Discrete token-based ASR}
In \cite{chang23b_interspeech}, ASR using discrete tokens as intermediate representations was proposed.
Discrete tokens, which represent each frame as a single integer,
offer high efficiency in terms of computation and memory usage.
In addition to this, this approach achieves higher recognition performance
compared to conventional methods with acoustic features such as log Mel-filterbanks.
Subsequently, research on discrete token-based ASR 
has been actively conducted \cite{chang2024exploring, yang2024towards,shi24h_interspeech, mousavi24_interspeech}.
However, these studies generally use the same dataset, thus same language,
for training the discrete tokens and the ASR model.
The evaluation is also conducted on the same in-domain data.
In this study, we focus on the use of discrete tokens trained on a different language
and examine the robustness of ASR against foreign-accented speech not included in the training data.


\section{Empirical analyses}

Figure \ref{fig:transcription} illustrates an overview of 
discrete token-based ASR for foreign-accented speech, aiming to simulate ISIB.
Based on the concept of \cite{onda24_interspeech}, shown in the section \ref{subsec:simulation},
discrete tokens are regarded as results of human perception.
When recognizing ``X-accented Y" speech,
by using discrete tokens trained on language X, 
the native language of the speaker,
the perceptive behaviors of listeners with the same native language X are simulated.
Then the recognition performance is expected to be improved through ISIB.
In this approach, ASR is trained only on native speech data of language Y, 
but the data are all converted into discrete tokens of language X.
The tokenization is also trained only on native speech data of language X.
Therefore, this approach aims to enhance recognition accuracy for ``X-accented Y" speech 
using only native speech data of languages X and Y. 
This makes the method particularly effective in scenarios 
where accented speech data is scarce.

\subsection{Experimental setup}
The experiments were conducted using ESPnet \cite{watanabe18_interspeech},
with the joint CTC/attention-based encoder-decoder
ASR model \cite{ctcaed}
following the configurations described in \cite{chang23b_interspeech,chang2024exploring}.
For input token sequences, subword modeling with byte pair encoding 
was not performed 
to facilitate comparison across different cluster sizes, and only 
deduplication (removing consecutive identical tokens) was applied.
We employed HuBERT-base \cite{hubert} as the SSL model for feature extraction,
following \cite{onda24_interspeech, lakhotiaetal2021generative}.
In all experiments, the ASR training was done for 50 epochs, 
and the model averaged over the top 3 epochs based on the validation set 
was used for evaluation.

First, experiments on Japanese-accented English
were conducted to be described in sections from \ref{sec:librispeech100} to \ref{sec:clusters}.
In this scenario, we compared the recognition performance for 
Japanese-accented English under two conditions:
when discretization via k-means clustering
was trained on English or Japanese.
For training k-means clusters, about 30 hours of native speech data were used for each language.
A random subset of the train-clean-100 of LibriSpeech \cite{libri} was 
used for English, and JVS \cite{takamichi2019jvs} was used for Japanese.
For evaluation, the ERJ corpus \cite{Minematsu2004DevelopmentOE} was used.
This corpus consists of English speech read by both Japanese learners
and native speakers of American English.
In our experiments, 460 phoneme-balanced sentences were used.
For Japanese speakers' speech, pronunciation assessment annotations 
are available in the corpus,
and we used the segmental scores for evaluation
as a measure of accent strength.

Lastly, in \ref{sec:mismatched}, experiments were extended to cover a wider range of accents.
Detailed settings are described in that section.

\subsection{LibriSpeech100 results}
\label{sec:librispeech100}

\begin{table}[tb]
	\centering
	\caption{ASR performance comparison: WER[\%] ($\downarrow$) (trained on LibriSpeech100)}
  \vspace*{-2mm}
	\label{tab:libri100}
    \resizebox{\columnwidth}{!}{
	\begin{tabular}{ccc!{\vrule}c!{\vrule}cc}
	  \toprule
		\multirow{2}{*}{cluster size} & \multirow{2}{*}{SSL} & \multirow{2}{*}{km} & LibriSpeech & \multicolumn{2}{c}{ERJ} \\
                                  &                      &                      &   test-\{clean, other\}                   & AE       & JE\_all/JE\_w10 \\
		\hline
    100 & en & en    & \textbf{10.1/24.0}     & \textbf{30.0}   & \textbf{91.1/101.9}    \\
          &    & jp   & 15.2/30.1 & 40.4 & 94.8/104.7 \\\cdashline{2-6}
          & jp & en   & 17.9/36.7 & 45.1 & 96.3/105.8 \\
          &     & jp  & 21.7/41.4 & 51.7 & 100.4/108.9 \\\hline
      500 & en & en  & \textbf{8.0/19.2}    & \textbf{23.6} & 83.8/95.9 \\
          &    & jp  & 9.1/21.1    & 27.4 & \textbf{82.7/95.2} \\\cdashline{2-6}
          & jp  & en  & 13.0/29.7&34.4 & 88.9/100.3 \\
          &     &jp   & 14.7/32.0 & 39.1 & 87.4/97.0 \\\hline
     2000 & en  & en  & \textbf{7.4/17.4} & \textbf{22.1} & 80.4/93.4 \\
           &     & jp & 8.1/18.9 & 23.9 & \textbf{79.5/92.0} \\\cdashline{2-6}
              & jp  & en &  10.7/26.2             & 30.8 & 86.5/98.0 \\
                &     & jp &   13.1/29.9          & 37.1 &  87.4/98.0\\
	  \bottomrule
	\end{tabular}
    }
  \vspace*{-5mm}
\end{table}

We first trained and evaluated ASR models 
using 
LibriSpeech100 dataset.
In this experiment, comparisons were made not only 
with the language used for k-means training 
but also the language for SSL pretraining and the cluster sizes.
We compared two pretrained HuBERT-base models:
one trained on English\footnote{\scriptsize \url{https://huggingface.co/facebook/hubert-base-ls960}}
and the other one trained on Japanese\footnote{\scriptsize \url{https://huggingface.co/rinna/japanese-hubert-base}}. 
In both cases, 
features were extracted from the 12th (last) layer of the model, 
and three cluster sizes were evaluated: 100, 500, and 2000.

The results are shown in Table \ref{tab:libri100}, 
where four conditions of the language for SSL pre-training and k-means training
(SSL, km) were tested for each cluster size.
The evaluation data includes the test sets of LibriSpeech,
as well as the ERJ corpus. 
For ERJ, American English (AE) and Japanese English (JE) were evaluated separately.
For JE, the results were shown
for all utterances (JE\_all) and for the subset from
the 10 worst speakers based on the segmental score annotations (JE\_w10),
who were supposed to have the strongest Japanese accent.

As for comparison between evaluation sets, in all cases,
in-domain LibriSpeech showed the highest accuracy, followed by ERJ (AE),
the out-of-domain native English data, and then ERJ (JE),
the out-of-domain Japanese-accented English.
Among ERJ (JE), the JE\_w10 subset showed lower accuracy than JE\_all.
Increasing the number of clusters
improved performance for all test sets.
This trend was also discussed in a previous study \cite{chang23b_interspeech},
and now it was confirmed that the same trend was observed even with
out-of-domain and accented speech data.
When using an SSL model pre-trained in Japanese, 
the accuracy deteriorated compared to the English-pretrained model
in all cases.
For SSL features, it is considered effective to use models trained on the target language
(in this case, English) regardless of accent.

When using the SSL model pretrained in English,
for native English (LibriSpeech and ERJ (AE)), the accuracy was higher
when the k-means clustering was trained on English data.
However, for Japanese-accented English (ERJ (JE)),
higher accuracy was observed when k-means was trained on Japanese,
when the cluster size was 500 or 2000.
This suggests that ISIB, where Japanese listeners have an advantage in understanding
Japanese-accented English, may have been replicated through the use of discrete tokens.
This trend was not observed when the cluster size was 100, probably because
the number of clusters was too small for effective speech recognition.
In fact, when the cluster size was 100, recognition accuracy even for native English speech 
also deteriorated significantly compared to when the cluster sizes were 500 or 2000.

\subsection{LibriSpeech960 results}
\label{sec:librispeech960}

Next, the experiment was extended to the LibriSpeech960 
to validate the occurrence of ISIB with a larger dataset.
The number of clusters was set to 2000, and 
the HuBERT-base pretrained in English was used
based on the results of the previous section.
In addition to the last layer, 
the cases with the 9th and 6th layers were also evaluated.

The results are shown in Table 2. 
In all cases, for the out-of-domain ERJ data,
AE were better recognized
when k-means clustering was done on English speech,
while JE 
showed better recognition accuracy
when k-means clusters were trained on Japanese.
This clearly exhibits the occurrence of ISIB.
Among different SSL layers,
there was a general trend where the difference between languages for k-means training ($\Delta$(en-jp))
became smaller as the layers approached the last layer. 
This is likely because upper layers have less acoustic information and more linguistic information \cite{hubertlayer}, 
allowing subtle pronunciation differences caused by accents to be less captured.

Table \ref{tab:example} shows an example of recognition results
of Japanese accented English, where 
/r/ was mispronounced like /l/ (a common error among Japanese speakers \cite{bradlow1997training, McKenzie03042015}).
The model trained with English k-means misrecognized the word ``appreciated'' as ``app\textit{li}ciated,''
while the model trained with Japanese k-means correctly recognized the word.
This suggests that clusters trained in English may distinguish the sounds of /r/ and /l/
more clearly, while those trained in Japanese are more tolerant of the difference,
making them more robust to Japanese accents.

\begin{table}[tb]
	\centering
	\caption{ASR performance comparison: WER[\%] ($\downarrow$) (trained on LibriSpeech960)}
  \vspace*{-2mm}
	\label{tab:libri960}
    \resizebox{\columnwidth}{!}{
	\begin{tabular}{ccc!{\vrule}c!{\vrule}cc}
	  \toprule
		\multirow{2}{*}{cluster size} & \multirow{2}{*}{SSL (layer)} & \multirow{2}{*}{km} & LibriSpeech & \multicolumn{2}{c}{ERJ} \\
                                  &                      &                      &   test-\{clean,other\}                   & AE       & JE\_all/JE\_w10 \\
		\hline
    2000  & en (12)  & en & 4.0/9.2 & \textbf{13.4} & 55.7/ 70.8 \\
           &     & jp & \textbf{3.8/9.2} & 14.3 &\textbf{53.3/68.0} \\\cdashline{3-6}
           &     & $\Delta$(en-jp) & 0.2/0.0 & -0.9 & 2.4/2.8 \\\hline
      & en (9) & en  & \textbf{3.5/8.8}    & \textbf{12.3} & 59.9/74.9 \\
          &    & jp  & 4.1/9.9   & 14.9 & \textbf{55.2/71.2} \\\cdashline{3-6}
          &     &  $\Delta$(en-jp)  &-0.6/-1.1& -2.6 & 4.7/3.7\\\hline
     & en (6) & en    & \textbf{3.7/9.4}     &  \textbf{13.5} &  56.9/71.3   \\
          &    & jp   & 4.8/11.0 & 16.8 & \textbf{54.0/67.2} \\\cdashline{3-6}
          &     & $\Delta$(en-jp)& -1.1/-1.6 & -3.3 & 2.9/4.1\\
	  \bottomrule
	\end{tabular}
    }
    \vspace*{-3mm}
\end{table}

\begin{table}[tb]
  \centering
  
  \caption{An example of recognition results}
  \vspace*{-2mm}
  \label{tab:example}
  \resizebox{\columnwidth}{!}{
  \begin{tabular}{ll}
    \toprule
    \multicolumn{2}{c}{Recognition result of IWA/M02/S2\_002} \\
    \hline
    GT & any contributions will be greatly appreciated \\
    en (9) & any contribution* will be greatly app\underline{\textit{li}}ciated \\
    jp (9) & any contribution* will be greatly app\textbf{re}ciated \\
    \bottomrule
  \end{tabular}
  }
  \vspace*{-5mm}
\end{table}

\subsection{The evaluation of clusters}
\label{sec:clusters}

The evaluation of the clustering quality
is typically conducted by comparing the discrete token sequences with the ground truth phoneme sequences,
such as phone normalized mutual information (PNMI) \cite{lakhotiaetal2021generative,hubert,chang23b_interspeech}.
However, for non-native speech, it is difficult to define accurate phoneme labels.
Therefore, in this study, we explore the following two metrics that do not
rely on the phonemic transcripts.\\
\noindent
\textbf{1) Quantization Error (QE):} The squared Euclidean distance between 
the original SSL features and the centroids of the clusters to which the features are assigned.
A smaller value indicates that the discretization is more suitable for the given data.\\
\noindent
\textbf{2) Mean Token Error Rate (MTER):} The average of token error rate (TER) 
computed across all possible pairs of multiple utterances of the same content.
TER is defined as the edit distance between two discrete token sequences 
generated from two utterances of the same content.
A lower MTER suggests that the tokenization is more robust to non-linguistic variations,
including accents.
Unlike word error rate (WER) or character error rate (CER),
MTER is calculated using multiple utterances as reference data for each sentence,
as ground truth token sequences cannot be defined.
In the ERJ corpus, the same sentence is read by 10 native speakers and 24 Japanese learners.
We used all the native speakers' utterances as reference data and
calculated TER for each utterance, then averaged the results.


The results are shown in Table \ref{tab:cluster}.
The evaluation was conducted using the ERJ corpus, and 
only the results for k-means models using English-pretrained SSL are shown.

Regarding QE, in most cases, 
the value was lower when k-means clustering was trained 
on English than on Japanese.
However, while when clusters are trained on English, AE showed lower QE
than JE,
the opposite trend was observed when k-means was trained with Japanese data.
As the proficiency of the speaker decreases as AE → JE\_all → JE\_w10,
the difference between the results of English and Japanese k-means
becomes smaller.
When using the 9th layer, QE with Japanese k-means got lower than with English k-means
for JE\_w10, indicating that Japanese k-means gives more adequate units for 
speech with stronger Japanese accents.

For MTER as well, k-means learned on English showed lower values
in most cases.
However, when the cluster size was 100 or 500,
the model trained on Japanese data showed lower MTER for JE\_w10.
This is likely because speakers with lower proficiency tend to have 
stronger accents, 
and Japanese sounds are more often substituted in the JE\_w10 utterances.
As a result, the distribution in the SSL feature space tends to 
fit better with the cluster centroids learned with Japanese data.
The reason why the same trend was not observed with the cluster size of 2000
is likely because as the number of clusters increased,
the case became more frequent where the same sentence 
spoken with different accents
was assigned to different clusters.
In the comparison between layers, the 9th layer showed 
lower MTER for AE than other layers.
We could say that this is 
related to the difference of ASR performances among layers.

\begin{table}[tb]
	\centering
	\caption{Cluster quality evaluation with ERJ corpus}
  \vspace*{-2mm}
	\label{tab:cluster}
    \resizebox{\columnwidth}{!}{
	\begin{tabular}{ccc!{\vrule}cc!{\vrule}cc}
	  \toprule
		\multirow{2}{*}{cluster size} & \multirow{2}{*}{SSL (layer)} & \multirow{2}{*}{km} & \multicolumn{2}{c!{\vrule}}{QE ($\downarrow$)} & \multicolumn{2}{c}{MTER [\%] ($\downarrow$)} \\
                                  &                      &                      & AE       & JE\_all/JE\_w10    & AE       & JE\_all/JE\_w10 \\
		\hline
	  100 & en (12) & en    &\textbf{79.1} &  \textbf{83.2/84.3}   & \textbf{39.4} &  \textbf{79.8}/94.7   \\
          &    & jp   & 94.9& 85.0/84.8& 50.8 & 83.2/\textbf{90.1} \\\hline
      500 & en (12)& en  & \textbf{56.9} &  \textbf{63.7/65.0} & \textbf{53.7} & \textbf{98.5}/114.1 \\
          &    & jp  & 78.8&  67.7/67.3  & 58.6 & 101.5/\textbf{112.3}\\\hline
     2000 & en (12) & en  & \textbf{47.1} & \textbf{54.9/56.3}& \textbf{66.7} & \textbf{108.7}/\textbf{121.1}  \\
           &     & jp & 66.8 &57.6/57.5 & 70.4 & 115.4/131.8  \\\hline
           & en (9) & en & \textbf{30.7} & \textbf{38.0}/39.8 & \textbf{53.7}  & \textbf{103.2/120.7} \\
           &     & jp & 43.5 & 39.1/\textbf{39.4} & 64.5 & 106.8/121.9 \\\hline
           & en (6) & en & \textbf{30.6} & \textbf{37.1/38.7} & \textbf{64.3} & \textbf{107.9/123.9} \\
           &     & jp & 41.4 & 38.9/40.1 &73.5 & 112.3/125.2 \\     
	  \bottomrule
	\end{tabular}
    }
    \vspace*{-5mm}
\end{table}

\subsection{Do mismatched ISIBs occur?}
\label{sec:mismatched}

Several studies reported the existence of ``mismatched" ISIB,
where non-native listeners have an advantage in understanding
non-native speech even when the native languages
of the speaker and listener are different \cite{bent2003interlanguage,yuan2010perception}.
While this is not always true for human listeners \cite{wang2015interlanguage,stibbard2006evidence},
we investigated whether such a phenomenon could be seen in discrete token-based ASR.

In this section, 
we examined more diverse cases where X-accented Y speech was recognized
using discrete tokens trained on either language X (``matched") or language Z (``mismatched").
We compared ASR performance on accented speech using k-means 
trained not only on English and Japanese but also on Chinese and Spanish. 
For k-means training in Chinese and Spanish, we used randomly selected
30-hour subsets 
from the AISHELL \cite{aishell} 
and MLS (Spanish) \cite{pratap20_interspeech}, respectively.
ASR training was done with LibriSpeech960, and the cluster size was set to 2000.
The 9th layer output of HuBERT-base was employed
since the layer showed the best performance for native English in the previous section.
For evaluation, in addition to ERJ, the L2-ARCTIC corpus \cite{zhao18b_interspeech} was used.
This corpus consists of English speech read by non-native speakers from 6 different native languages,
including Chinese and Spanish.

The results are shown in Table \ref{tab:multi}.
For each test set, the best and worst results are shown in bold and underlined, respectively.
For all the ``matched" case, 
as shown in the diagonal elements of the first four columns,
the recognition accuracy was
highest when k-means clustering was trained on the same language as the speaker's native language.
This suggests that the ``matched" ISIB was also observed in
Chinese and Spanish-accented English.
As for the ``mismatched" cases, 
when k-means was trained on Chinese or Spanish,
the recognition accuracy for all foreign accents 
was higher than when trained on English.
In most cases,
k-means trained on English showed the worst performance.
This suggests that the ``mismatched'' ISIB occurs in 
the discrete token-based ASR.
We may say that this approach is effective even when the speakers' native language is
unknown if only they are non-native speakers.

For Arabic and Hindi-accented English, the recognition accuracy was the highest
when k-means clustering was trained on Spanish.
For Korean and Vietnamese-accented English, 
when using Chinese-trained k-means showed the best performance.
This suggests that the ``mismatched'' ISIB on the discrete token-based ASR
is more effective
when the 
accents of the speaker and the listener are closer.
In fact, a study on accent classification 
showed some confusion between 
Arabic and Spanish, and between Vietnamese and Chinese accents \cite{yang23v_interspeech}. 
This indicates that this method can be applied even when the speaker's native language is
low-resource and it is difficult to obtain enough speech data for k-means training.
By selecting a more high-resource language the accent of which is
similar to the speaker's accent, the recognition performance may be further improved.

\begin{table}[tb]
	\centering
	\caption{ASR results for more diverse accents. Cluster size: 2000, SSL: en (9). WER[\%] ($\downarrow$) (trained on LibriSpeech960)}
  \vspace*{-2mm}
  \hspace*{-5mm}
	\label{tab:multi}
    \resizebox{1.1\columnwidth}{!}{
	\begin{tabular}{c!{\vrule}cc!{\vrule}ccccccc}
	  \toprule
		 \multirow{2}{*}{km} & \multicolumn{2}{c!{\vrule}}{ERJ} & \multicolumn{6}{c}{L2-ARCTIC} \\
                         & AE       & JE\_all/JE\_w10 & Chinese & Spanish & Arabic & Hindi & Korean & Vietnamese \\\hline 
    en & \textbf{12.3} & \underline{59.9/74.9}  & 29.8 & \underline{23.4} & \underline{23.5} & \underline{20.0} & \underline{19.6} & 35.9 \\
    jp & \underline{14.9} & \textbf{55.2/71.2}  & \underline{30.1} & 23.3 & 23.4 & 19.8 & 19.2 & \underline{36.9} \\
    ch & 13.7 & 57.3/71.6 & \textbf{28.6} & 22.4 & 22.4 & 19.6 & \textbf{18.6} & \textbf{34.9} \\
    sp & 14.4 & 57.1/71.4  & 29.3 & \textbf{21.8} & \textbf{22.1} & \textbf{19.3} & 18.9 & 35.3 \\
	  \bottomrule
	\end{tabular}
    }
    \vspace*{-5mm}
\end{table}

\section{Conclusions}
In this study, we conducted comparative experiments on discrete token-based ASR
for foreign-accented speech varying the language used for k-means clustering.
The results showed that training k-means on the speaker's native language
led to the best recognition performance
among the tested languages, including the spoken language.
This shows the occurrence of ISIB in discrete token-based ASR,
and supports the idea suggested in \cite{onda24_interspeech} more strongly
that discrete tokens simulate the human perception of speech.
Since our approach of replicating ISIB relies only on native speech data,
it can be applied 
even if the accented speech data is unavailable.
The fact that ``mismatched'' ISIB was also observed in discrete token-based ASR
shows further applicability of this approach, even when the speaker's native language is unknown or low-resource.
Future work includes exploring more advanced simulations of speech perception
by multilingual listeners and
improving recognition performance through them
while using only native speech data.


\section{Acknowledgements} 
This work was supported by AIST KAKUSEI project (FY2024).

\bibliographystyle{IEEEtran}
\bibliography{references}

\end{document}